\documentstyle[epsf]{elsart}
\newcommand{\D}{d}

\begin{document}
\begin{frontmatter}
%
\title{Numerical Relativity As A Tool For Computational Astrophysics}
%
\author[AUT1]{Edward Seidel
}
%
\author[AUT2]{Wai-Mo Suen
}
%
\address[AUT1]{Max-Planck-Institut f\"ur Gravitationsphysik \\
Albert-Einstein-Institut \\
Schlaatzweg 1, D-14473, Potsdam, Germany\\
{\rm and} \\
University of Illinois \\
NCSA and Departments of Physics and Astronomy \\
Champaign, IL, 61820, USA
}
%
\address[AUT2]{McDonnell Center for the Space Sciences \\
Physics Department \\
Washington University, St. Louis, MO63130, U.S.A \\
{\rm and} \\
Physics Department \\
Chinese University of Hong Kong \\
Shatin, Hong Kong
}
%
\begin{abstract}

The astrophysics of compact objects, which requires Einstein's theory of
general relativity for understanding phenomena such as black holes and neutron
stars, is attracting increasing attention.  In general relativity,
gravity is governed by an extremely complex set of coupled, nonlinear,
hyperbolic-elliptic partial differential equations.  The largest
parallel supercomputers are finally approaching the speed and memory
required to solve the complete set of Einstein's equations for the
first time since they were written over 80 years ago, allowing one to
attempt full 3D simulations of such exciting events as colliding black
holes and neutron stars.  In this paper we review the computational
effort in this direction, and discuss a new 3D multi-purpose
parallel code called ``Cactus'' for general relativistic astrophysics.
Directions for further work are indicated where appropriate.
\end{abstract}
\end{frontmatter}

\section{Overview}

The Einstein equations for the structure of spacetime were first
published in 1916 when Einstein introduced his famous general theory
of relativity.  This theory of gravity has remained essentially
unchanged since its discovery, and it provides the underpinnings of
modern theories of astrophysics and cosmology. The theory is essential
in describing phenomena such as black holes, compact objects,
supernovae, and the formation of structure in the Universe.
Unfortunately, the equations are a set of highly complex, coupled,
nonlinear partial differential equations involving 10 functions of 4
independent variables.  They are among the most complicated equations
in mathematical physics.  For this reason, in spite of more than 80
years of intense study, the solution space to the full set of
equations is essentially unknown.  Most of what we know about this
fundamental theory of physics has been gleaned from linearized
solutions, highly idealized solutions possessing a high degree of
symmetry (e.g., static, or spherically or axially symmetric), or from
perturbations of these solutions.

Over the last 30 years a growing research area, called Numerical
Relativity, has developed, where computers are employed to construct
numerical solutions to these equations.  Although much has been
learned through this approach, progress has been slow due to the
complexity of the equations and inadequate computational power.  For
example, an important astrophysical application is the 3D spiraling
coalescence of two black holes (BH) or neutron stars (NS), which will generate
strong sources of gravitational waves.  As has been emphasized by
Flanagan and Hughes, one of the best candidates for early detection by
the laser interferometer network is increasingly considered to be BH
mergers\cite{Flanagan97b,Flanagan97a}.  The imminent arrival of data
from the long awaited gravitational wave interferometers (see,
e.g., Ref.~\cite{Flanagan97b} and references therein) has provided a
sense of urgency in understanding these strong sources of
gravitational waves.  Such understanding can be obtained only through
large scale computer simulations using the full machinery of numerical
relativity.

Furthermore, the gravitational wave signals are likely to be so weak
by the time they reach the detectors that reliable detection may be
difficult without prior knowledge of the merger waveform.  These
signals can be properly interpreted, or perhaps even detected, only
with a detailed comparison between the observational data and a set of
theoretically determined ``waveform templates''.  In most cases, these
waveform templates needed for gravitational wave data analysis have to
be generated by large scale computer simulations, adding to the
urgency of developing numerical relativity.  However, a realistic 3D
simulation based on the full Einstein equations is a highly
non-trivial task---based on axisymmetric black hole calculations
performed during late 1980's and algorithms available at the
time---one can estimate the time required for a reasonably accurate 3D
simulation of, say, the coalescence of a compact object binary, to be
at least 100,000 Cray Y-MP hours!

But there is good reason for optimism that such problems can be solved
within the next decade.  Scalable parallel computers, and efficient
algorithms that exploit them, are quickly revolutionizing
computational science, and numerical relativity is a great beneficiary
of these developments.  Over the last years the community has
developed 3D codes designed to solve the complete set of Einstein
equations that run very efficiently on large scale parallel computers.
We will describe below one such code, called ``Cactus'', that has
achieved 142 GFlops on a 1024 node Cray T3E-1200, which is more than
2000 times faster than 2D codes of a few years ago running on a Cray
Y-MP (which also had only about 0.5\% the memory capacity of the large
T3E).  Such machines are expected to scale up rapidly as faster
processors are connected together in even higher numbers, achieving
Teraflop performance on real applications in a few years.

Numerical relativity requires not only large computers and efficient
codes, but also a wide variety of numerical algorithms for evolving
and analyzing the solution.  Because of this richness and complexity
of the equations, and the interesting applications to problems such as
black holes and neutron stars, natural collaborations have developed
between applied mathematicians, physicists, astrophysicists, and
computational scientists in the development of a single code to attack
these problems.  There are various large scale collaborative effort in
recent years in this direction, including the NSF Black Hole Grand
Challenge Project (recently concluded), the NASA Neutron Star Grand
Challenge Project and the NCSA/Potsdam/Wash U numerical relativity
collaboration.

We introduce in this paper a code called ``Cactus'', which is
developed by the NCSA/Potsdam/Wash U collaboration, and is employed in
the NASA Neutron Star Grand Challenge Project. 
We will describe some of the algorithms and capabilities of this code
in this paper.  In the next sections we will first give a
brief description of the numerical formulation of the theory of
general relativity, and discuss particular difficulties associated
with numerical relativity.  The discussion will necessarily be brief.
Examples are mostly drawn from work carried out by our
NCSA/Potsdam/Wash U numerical relativity collaboration.  We also
provide URL addresses for web pages containing graphics and movies of
some of our results.

To conclude this brief introduction, a statement of where we stand in 
terms of simulating general relativistic compact objects is in order.  
The NSF black hole grand challenge project and related work achieved 
long term stable evolution of single black hole spacetimes under 
certain conditions~\cite{Daues96a,Cook97a,Gomez98a}, but there is 
still a long way to go before the spiraling coalescence can be 
computed.  The presently on-going NASA neutron star grand challenge 
project recently succeeded in evolving grazing collision of two 
neutron stars using the full Einstein-relativistic hydrodynamic system 
of equations, with a simple equation of state.  While the inspiral 
coalescences of two neutron stars is {\it not} a stated goal of the 
NASA project, we expect to be able to carry out preliminary studies of 
the inspiral coalescences in the next few years.  The final goal of a 
full solution of the problem including radiation transport and 
magneto-hydrodynamics for comparison between numerical simulations and 
observations in gravitational wave astronomy (waveform templates) and 
high energy astronomy ($gamma$ ray burst) will take many more years, 
hopefully building on the effort described in this paper.  The 
Nakamura group also reports preliminary success in evolving several 
orbits with a fully relativistic GR-hydro code~\cite{Nakamura99a}.

\section{Einstein Equations for Relativity}

The generality and complexity of the Einstein equations make them an
excellent and fertile testing ground for a variety of broadly
significant computing issues.  They form a system of dozens of
coupled, nonlinear equations, with thousands of terms, of mixed
hyperbolic-elliptic type, and even undefined types, depending on
coordinate conditions.  This rich and general structure of the
equations implies that the techniques developed to solve our problems
will be immediately applicable to a large family of diverse scientific
applications.

The system of equations breaks up naturally into a set of constraint
equations, which are elliptic in nature, evolution equations, which
are ``hyperbolic'' in nature (more on this below), and gauge
equations, which can be chosen arbitrarily (often leading to more
elliptic equations).  The evolution equations guarantee (mathematically)
that the elliptic constraints are satisfied at all times provided the
initial data satisfied them.  This implies that the initial data are
not freely specifiable.  Moreover, although the constraints are
satisfied mathematically during evolution, it will not be so numerically.
These problems are each discussed in turn below.  First, however, we
point out that a much simpler theory, familiar to many, has all of
these same features.  Maxwell's equations describing electromagnetic
radiation have: (a) elliptic constraint equations, demanding that in
vacuum the divergence of the electric and magnetic fields vanish at
all times; (b) evolution equations, determining the time development
of these fields, given suitable initial data that satisfies the elliptic
constrain equations; and (c) gauge conditions that can be applied
freely to certain variables in the theory, such as some components of
the vector potential.  Some choices of vector potential lead to
hyperbolic evolution equations for the system, and some do not.  We
will find all of these features present in the much more complicated
Einstein equations, so it is useful to keep Maxwell's equations in
mind when reading the next sections.

In the standard 3+1 ADM approach to general relativity,\cite{York79},
the basic building block of the theory---the spacetime metric---is written
in the form
\begin{equation}
 \D s^2   = -(\alpha^{2} -\beta ^{a}\beta _{a}) \D t^2
 + 2 \beta _{a}  \D x^{a} \D t
 +\gamma_{ab} \D x^{a} \D x^{b}\; ,
\end{equation}
using geometrized units such that the gravitational constant $G$ and
the speed of light $c$ are both equal to unity. Throughout this paper,
we use Latin indices to label spatial coordinates, running from 1 to
3.  The ten functions $(\alpha, \beta ^{a}, \gamma_{ab})$ are
functions of the spatial coordinates $x^{a}$ and time $t$.  Indices
are raised and lowered by the ``spatial 3--metric'' $\gamma_{ab}$.
Notice that the geometry on a 3D spacelike
hypersurface of constant time (i.e., $dt=0$) is determined by
$\gamma_{ab}$.  As we will see below, the Einstein equations control
the evolution in time of this 3D geometry described by $\gamma_{ab}$,
given appropriate initial conditions.  The lapse function $\alpha$ and
the shift vector $\beta^{a}$ determine how the slices are threaded by
the spatial coordinates.  Together, $\alpha$ and $\beta^a$ represent
the coordinate degrees of freedom inherent in the covariant
formulation of Einstein's equations, and can therefore be chosen, in
some sense, ``freely'', as discussed below.

This formulation of the equations assumes that one begins with an
everywhere spacelike slice of spacetime, that should be evolved
forward in time.  Due to limited space, we will not discuss promising
alternate treatments, based on either characteristic, or null
foliations of spacetime\cite{Bishop98a}, or on asymptotically null
slices of
spacetime\cite{Friedrich81a,Friedrich81b,Friedrich96,Huebner96}.

\subsection{Constraint Equations}
\label{constraints}

The constraints can be considered as the relativistic generalization
of the Poisson equation of Newtonian gravity, but instead of a single
linear elliptic equation there are now four, coupled, highly nonlinear
elliptic equations, known as the hamiltonian and momentum constraints.
Under certain conditions, the equations decouple and can be solved
independently and more easily, and this is how they have been usually
treated.  Recently, techniques have been developed that allow one to
solve the constraints in a more general setting, without making
restrictive assumptions that lead to
decoupling\cite{Bernstein97x,Laguna97x,Thornburg98,Miller98x}.  In
such a system the four constraint equations are solved simultaneously.
This may prove useful in generating new classes of initial data.
However, at present there is no satisfactory algorithm for
controlling the physics content of the data generated.  The major
remaining work in this direction is to develop a scheme that is
capable of constructing the initial data that describe a {\it given}
physical system.  That is, although we have schemes available to
solve many variations on the initial value problem, it is difficult
to specify in advance, for example, what are the precise spins and momenta of
two black holes in orbit, or even if the hole {\em are} in orbit.  This
can generally only be determined after the equations have been solved
and analyzed.

The elliptic operators for these equations are usually symmetric, but
they are otherwise the most general type, with all first and mixed second
derivative terms present.  The boundary conditions, which can break
the symmetry, are usually linear conditions that involve derivatives
of the fields being solved.  In any case, once the initial value
equations have been solved, initial data for the evolution problem
result.

We illustrate the central idea of constructing initial data with
vacuum spacetimes for simplicity.  The application of the algorithm
presented here to a general spacetime with matter source is currently
routine in numerical relativity.  The full 4D Einstein equations can
be decomposed into six evolution equations and four constraint
equations.  The constraints\index{constraints!in relativity} may be
subdivided, in turn, into one Hamiltonian (or energy) constraint
equation,
\begin{equation}
\label{Hamiltonian constraint}
R+({\mathrm tr}\, K)^{2}-K^{ab}K_{ab}=0\; ,
\end{equation}
and three momentum constraint equations (or one vector equation),
\begin{equation}
\label{momentum constraint}
D_{b}(K^{ab}-\gamma^{ab}{\mathrm tr}\, K)=0\; .
\end{equation}
In these equations $K_{ab}$ is the extrinsic curvature of the slice,
related to the
time derivative of $\gamma_{ab}$ by
\begin{equation}
K_{ab} = -\frac{1}{2 \alpha}
(\partial_t \gamma_{ab} - D_a \beta_b -D_b \beta_a)\; .
\label{kdef}
\end{equation}
Here we have introduced the 3D spatial covariant derivative operator
$D_a$ associated with the 3--metric $\gamma_{ab}$ (i.e. $D_a
\gamma_{bc}=0$), and the 3D scalar curvature $R$ computed from $\gamma_{ab}$.
These four constraint equations can be used to
determine initial data for $\gamma_{ab}$ and $K_{ab}$, which are to be
evolved with the evolution equations to be discussed below.
These equations (2,3) are referred to as constraints because, as in the case
of electrodynamics, they contain no time derivatives of the
fundamental fields $\gamma_{ab}$ and $K_{ab}$, and hence do not
propagate the solution in time.

Next, we will sketch the standard method for obtaining a solution to
these constraint equations.  We follow York and coworkers (e.g.,
\cite{York89}) by writing the 3--metric and extrinsic curvature in
``conformal form'', and also make use of the simplifying assumption
$\mathrm{tr}K = 0$ which causes the Hamiltonian and momentum
constraints to completely decouple (note that actually the equations
decoupled with $\mathrm{tr}K = {\rm const.}$ but we will discuss only
the simplest case here).  We write
\begin{equation}
\label{conformal form}
\gamma_{ab}=\Psi^{4}\hat{\gamma}_{ab}, K_{ab} = \Psi^{-2}\hat{K}_{ab}\; ,
\end{equation}
where $\hat{\gamma}_{ab}$ and the transverse tracefree part of
$\hat{K}_{ab}$ is regarded as given, i.e., chosen to represent the
physical system that we want to study.  Under the
conformal transformation, with $\mathrm{tr}K = 0$ we find that the
momentum constraint becomes
\begin{equation}
\hat{D}_b \hat{K}^{ab} = 0\; ,
\label{eq10}
\end{equation}
where $\hat{D}_a$ is the 3D covariant derivative associated with
$\hat{\gamma}_{ab}$ (i.e., $\hat{D}_a \hat{\gamma}_{ab}=0$).  In vacuum,
black hole spacetimes $\hat{K}_{ab}$ can often be
solved analytically.
For more details on how to solve the momentum constraints in complicated
situations, please see \cite{York79,Cook93,Nakamura89}.

The remaining unknown function $\Psi$,
must satisfy the Hamiltonian constraint.
The conformal transformation of the scalar curvature is
\begin{equation}
R=\Psi^{-4}\hat{R}-8\Psi^{-5}\hat{\Delta}\Psi\; ,
\end{equation}
where $\hat{\Delta} = \hat{\gamma}^{ab} \hat{D}_a \hat{D}_b$ and $\hat{R}$ is
the scalar curvature of the known metric $\hat{\gamma}_{ab}$.
Plugging this back in to the Hamiltonian constraint and
dividing through by $-8 \Psi^{-5}$, we obtain
\begin{equation}
\hat\Delta\Psi-\frac{1}{8}\Psi \hat{R}+\frac{1}{8}
\Psi^{-7} \left( \hat{K}_{ab} \hat{K}^{ab}\right) = 0\; ,
\label{psieq}
\end{equation}
an elliptic equation for the conformal factor $\Psi$.

To summarize, one first specifies $\hat{\gamma}_{ab}$ and the 
transverse tracefree part of $\hat{K}_{ab}$ ``at will'', choosing them 
to be something ``closest'' to the spacetime one wants to study.  Then 
one solves (\ref{eq10}) for the conformal extrinsic curvature 
$\hat{K}_{ab}$.  Finally, (\ref{psieq}) is solved for the conformal 
factor $\Psi$, so the full solution $\gamma_{ab}$ and $K_{ab}$ can be 
reconstructed.  In this process the elliptic equations are solved by 
standard techniques, e.g., the conjugate gradient~\cite{Ashby90} or 
multigrid methods~\cite{Cook89}.  In situations where there is a black 
hole singularity, there could be added complications in solving the 
elliptic equations, and special treatments would have to be 
introduced, e.g., the ``puncture'' treatment of \cite{Brandt97b}, or 
employing an ``isometry'' operation to provide boundary conditions on 
black hole throats, ensuring identical spatial geometries inside and 
outside the throat (see, e.g.,~\cite{Cook91,Cook93}, or~\cite{Seidel96a} for 
more details).

While this is a well established process for generating an initial
data set for numerical study, there is a fundamental difficulty in
using this approach to generate initial data corresponding to a
physical system one wants to evolve, e.g., a coalescing binary system.
It is not clear how to choose the ``closest'' $\hat{\gamma}_{ab}$, and
the corresponding free components in $\hat{K}_{ab}$, so that the
resulting $\gamma_{ab}$ and $K_{ab}$ represents the inspiraling
system at its late stage of inspiral.  This late stage is the
so-called ``intermediate challenge problem'' of binary black
holes~\cite{Brady98a}, an area of much current interest.

\subsection{Evolution Equations}
\label{evolution}
\subsubsection{The standard evolution system}

With the initial data $\gamma_{ab}$ and $K_{ab}$ specified, we now consider
their evolution in time.
There are six evolution equations for the 3--metric $\gamma_{ab}$ that
are second order in time, resulting from projections of the full
4D Einstein equations onto the 3D spacelike slice \cite{York79}.
These are most often written as a first-order-in-time system of twelve
evolution equations, usually referred to as the ``ADM'' evolution
system~\cite{Arnowitt62,York79}:
\begin{eqnarray}
\partial_{t}\gamma_{ab}
      & = & -2\alpha K_{ab}+
            D_{a}\beta_{b}+D_{b}\beta_{a}\;
\label{metric evolution}\\
\partial_{t}K_{ab}
    & = & -D_{a}D_{b}\alpha+\alpha \left[
          R_{ab}+\mbox({tr}K)K_{ab}-2K_{ac}K^{c}{}_{b} \right] \nonumber\\
    &   & +\beta^{c}D_{c}K_{ab}+
          K_{ac}D_{b}\beta^{c}+K_{cb}D_{a}\beta^{c}\; .\label{excurv evolution}
\end{eqnarray}
Here $R_{ab}$ is the Ricci tensor of the 3D spacelike slice labeled by
a constant value of $t$.  Note that these are quantities defined only
on a $t=const$ hypersurface, and require only the 3--metric
$\gamma_{ab}$ in their construction.  Do not confuse them with the
conventional 4D objects!  The complete set of Einstein equations are
contained in constraint equations (\ref{Hamiltonian constraint}),
(\ref{momentum constraint}) and the evolution equations (\ref{excurv
evolution}), (\ref{metric evolution}).  Note that (\ref{metric
evolution}) is simply the definition of the extrinsic curvature
$K_{ab}$ (\ref{kdef}).  These equations are analogous to the evolution
equations for the electric and magnetic fields of electrodynamics.
Given the ``lapse'' $\alpha$ and ``shift'' $\beta^a$, discussed below,
they allow one to advance the system forward in time.

\subsubsection{Hyperbolic evolution systems}

The evolution equations (\ref{excurv evolution}), (\ref{metric
evolution}) have been presented in the ``standard ADM form'', which
has served numerical relativity well over the last few decades.
However, the equations are enormously complicated; the complication is
hidden in the definition of the curvature tensor $R_{ab}$ and the
covariant differentiation operator $D_a$.  In particular, although
they describe physical information propagating with a finite speed,
the system does not form a hyperbolic system, and is not necessarily
the best for numerical evolution.  Other fields of physics, in
particular hydrodynamics, have developed very mature numerical methods
that are specially designed to treat the well studied flux
conservative, hyperbolic system of balance laws having the form
\begin{equation}
\partial_t {\bf u} + \partial_k F^k_-{\bf u} = S_-{\bf u}
\label{3Dbalance}
\end{equation}
where the vector ${\bf u}$ displays the set of variables and both
``fluxes'' $F^k$ and ``sources'' $S$ are vector valued functions.  In
hydrodynamic systems, it often turns out that the characteristic matrix
$\partial F / \partial
{\bf u} $ projected into any spacelike direction can often be diagonalized,
so that fields with definite propagation speeds can be identified
(the eigenvectors and the eigenvalues of the projected characteristic matrix).
One important point is that in (\ref{3Dbalance}) all spatial derivatives
are contained in the flux
terms, with the source terms in the equations containing no
derivatives of the eigenfields.  All of these features can be
exploited in numerical finite difference schemes that treat each term
in an appropriate way to preserve important physical characteristics
of the solution.

Amazingly, the complete set of Einstein equations can also be put in
this ``simple'' form (the source terms still contain thousands of
terms however).  Building on earlier work by Choquet-Bruhat and
Ruggeri\cite{Choquet83}, Bona and Mass\'o began to study this problem
in the late 1980's, and by 1992 they had developed a hyperbolic system
for the Einstein equations with a certain specific gauge
choice\cite{Bona92} (see below).  Here by hyperbolic, we mean simply
that the projected characteristic matrix has a complete set of
eigenfields with real eigenvalues.  This work was generalized recently
to apply to a large family of gauge choices\cite{Bona94b,Bona97a}.
The Bona-Mass\'o system of equations is available in the 3D ``Cactus''
code~\cite{Bona98b,Alcubierre98a}, as is the standard ADM system, 
where both are tested and compared on a number of spacetimes.

The Bona-Mass\'o system is now one among many hyperbolic systems, as 
other independent hyperbolic formulations of Einstein's equations were 
developed\cite{Fritelli94,Choquet95,Abrahams95a,Fritelli95,MVP95,Abrahams97b} 
at about the same time as Ref.~\cite{Bona94a}.  Among these other 
formulations only the one originally devised in 
Ref.~\cite{Abrahams95a} has been applied to spacetimes containing 
black holes\cite{Scheel97}, although still only in the spherically 
symmetry 1D case (a 3D version is under 
development\cite{CookScheelPrivateComm}.)  Hence, of the many 
hyperbolic variants, only the Bona-Mass\'o family and the formulations 
of York and co-workers have been tested in any detail in 3D numerical 
codes.  Notably among the differences in the formulations, the 
Bona-Mass\'o and Fritelli families contain terms equivalent to second 
time derivatives of the three metric $\gamma_{ab}$, while many other 
formulations go to a higher time derivative to achieve hyperbolicity.  
Another comment worth making is that for harmonic slicing, both the 
Bona-Mass\'o and York families have characteristic speeds of either 
zero, or light speed.  For maximal slicing, they both reduce to a 
coupled elliptic-hyperbolic system.  The Bona-Mass\'o system (at least) 
also allows for an additional family of explicit algebraic slicings, 
with the lapse proportional to an explicit function of the determinant 
of the three--metric, and in those cases one can also identify gauge 
speeds which can be different from light speed (harmonic slicing is 
one example of this family where the gauge speed corresponds to light 
speed).  Some of these slicings, such as ``1+log''~\cite{Anninos94c}, 
have been found to be very useful in 3D numerical evolutions.  This 
information about the speed of gauge and physical propagation can be 
very helpful in understanding the system, and can also be useful in 
developing numerical methods.  Only extensive numerical studies will 
tell if the various hyperbolic formulations live up to their promise.

Reula has recently reviewed, from the mathematical point of view, most
of the recent hyperbolic formulations of the Einstein
equations\cite{Reula98a} (This article, in the online journal ``Living
Reviews in Relativity'', will be periodically updated).  It is
important to realize that the mathematical relativity field has been
interested in hyperbolic formulations of the Einstein equations for
many years and some systems that could have been suitable for
numerical relativity were already published in the
1980's\cite{Choquet83,Friedrich85}.  However, these developments were
generally not recognized by the numerical relativity community until
recently.

\subsubsection{Numerical techniques for the evolution equations}
Most of what has been attempted in numerical relativity evolution
schemes is built on explicit finite difference schemes.  Implicit and
iterative evolution schemes have been occasionally attempted, but the
extra cost associated has made them less popular.  We now describe the
basic approach that has been tried for both the standard ADM
formulation and more recent hyperbolic formulations of the equations.

\paragraph{ADM evolutions}

The ADM system of evolution equations is often solved using some
variation of the leapfrog method, similar to that described in
have been used successfully.  The most extensively tested is the
``staggered leapfrog'', detailed in axisymmetric cases in
Ref.~\cite{Bernstein93b} and in 3D in Ref.~\cite{Anninos94c}, but
other successful versions include full leapfrog implementations used
in 3D by \cite{Bruegmann96} and \cite{Bona98b}.  For the ADM system,
the basic strategy is to use centered spatial differences everywhere,
march forward according to some explicit time scheme, and hope for the
best!  Generally, this technique has worked surprisingly well until
large gradients are encountered, at which time the methods often break
down.  The problem is that the equations in this ADM form are
difficult to analyze, and hence ad hoc numerical schemes are often
tried without detailed knowledge of how to treat specific terms in the
equations, or how to treat instabilities when they arise.  A
recent development is that of the ``deloused'' leapfrog, which amounts
to filtering the solution\cite{New98}.  Also recently, the iterative
Crank-Nicholson scheme has been found effective in suppressing
some instabilities that occur~\cite{Huq98x}.

\paragraph{Hyperbolic evolutions}
The hyperbolic formulations are on a much firmer footing numerically
than the ADM formulation, as the equations are in a much simpler form
that has been studied for many years in computational fluid dynamics.
However, the application of such methods to relativity is quite new,
and hence the experience with such methods in this community is
relatively limited.  Furthermore, the treatment of the highly nonlinear
source terms that arise in relativity is very much unexplored, and the
source terms in Einstein's equations are much more complicated than
those in hydrodynamics.

A standard technique for equations having flux conservative form is to
split Eq.~(\ref{3Dbalance}) into two separate processes.  The
transport part is given by the flux terms
\begin{equation}
  \partial_t {\bf u} + \partial_k F^k_-{\bf u} = 0 \;\; .
\label{3Dtransport}
\end{equation}
The source contribution is given by the following system of {\em
ordinary} differential equations
\begin{equation}
  \partial_t {\bf u} = S_-{\bf u} \;\; .
\label{3Dsources}
\end{equation}
Numerically, this splitting is performed by a combination of both flux
and source operators. Denoting by $E(\Delta t)$ the numerical evolution
operator for system (\ref{3Dbalance}) in a single timestep, we
implement the following combination sequence of subevolution steps:
\begin{equation}
  E(\Delta t) = S(\Delta t/2)\;T(\Delta t)\;S(\Delta t/2)
\label{splitting}
\end{equation}
where $T$, $S$ are the numerical evolution operators for systems
(\ref{3Dtransport}) and (\ref{3Dsources}), respectively.  This is
known as ``Strang splitting''~\cite{Press86}.  As long as both
operators $T$ and $S$ are second order accurate in $\Delta t$, the
overall step of operator $E$ is also second order accurate in time.

This choice of splitting allows easy implementation of different
numerical treatments of the principal part of the system without
having to worry about the sources of the equations.  Additionally,
there are numerous computational advantages to this technique, as
discussed in \cite{Clune98a}.

The sources can be updated using a variety of ODE integrators, and in
``Cactus'' the usual technique involves second order predictor-corrector
methods.  Higher order methods for source
integration can be easily implemented, but this will not improve the
overall order of accuracy.  However, in special cases where the
evolution is largely source driven\cite{Masso92}, it may be important
to use higher order source operators, and this method allows such
generalizations.  The details can be found in Ref.~\cite{Bona98b}.

The implementation of numerical methods for the flux operator is much
more involved, and there are many possibilities, ranging from standard
choices to advanced shock capturing
methods\cite{Leveque92,Bona96a,Bona97a}.  Among standard methods, the
MacCormack method, which has proven to be very robust in the
computational fluid dynamics field (see, e.g., Ref.~\cite{Yee88} and
references therein), and a directionally split Lax-Wendroff method
have been implemented and tested extensively in ``Cactus''.
These schemes are fully second order in space and time.  Shock
capturing methods have been shown to work extremely well in 1D
problems in numerical relativity~\cite{Bona94b,Bona96a}, but their
application in 3D is an active research area full of promise, but as
yet, unfulfilled.  The details of these methods, as they are applied
to the Bona-Mass\'o formulation of the equations, can be found in
Refs.~\cite{Bona96a,Bona98b}.

\subsubsection{The Role of Constraints}
If the constraints are satisfied on the initial hypersurface, the 
evolution equations then guarantee that they remain satisfied on all 
subsequent hypersurfaces.  Thus once the initial value problem has 
been solved, one may advance the solution forward in time by using 
only the evolution equations.  This is the same situation encountered 
in electrodynamics as discussed above.  However, in a numerical 
solution, the constraints\index{constraints!in relativity} will be 
violated at some level due to numerical error.  They hence provide 
useful indicators for the accuracy of the numerical spacetimes 
generated.  Traditional alternatives to this approach, which is often 
referred to as ``free evolution'', involve solving some or all of the 
constraint equations on each slice for certain metric and extrinsic 
curvature components, and then simply monitoring the ``left over'' 
evolution equations.  This issue is discussed further by Choptuik in 
\cite{Choptuik91}, and in detail for the Schwarzschild spacetime in 
\cite{Bernstein89}.  New approaches to this problem of constraint vs.  
evolution equations are currently being pursued by 
Lee~\cite{Lee93,Lee94a}, among others.  This approach is to advance 
the system forward using the evolution equations, and then adjust the 
variables slightly so that the constraints are satisfied (to some 
tolerance), i.e., the solution is projected onto the constraint 
surface.  Because there are many variables that go into the 
constraints, there is not a unique way to decide which ones to adjust 
and by how much.  But one can compute the ``minimum'' perturbation to 
the system, which corresponds to projecting to the {\em closest} point 
on the constraint surface.  Other approaches, similar in spirit to 
each other, have been suggested by Detweiler~\cite{Detweiler87} and 
Brodbeck et al~\cite{Brodbeck98}.  The Detweiler approach restricts 
the numerical evolution to the constraint surface by adding terms to 
the evolution equations (\ref{metric evolution}), (\ref{excurv 
evolution}) terms which are proportional to the constraints.  
Numerical tests of the scheme using gravitational wave spacetimes have 
recently been carried out, showing promising results~\cite{Lai98}.

\subsubsection{Gauge Conditions}
\label{gauge}

Kinematic conditions for the lapse function $\alpha$ and shift vector
$\beta^i$ have to be specified for the evolution equations
(\ref{metric evolution}), and (\ref{excurv evolution}).  With
$\gamma_{ab}$ and $K_{ab}$ satisfying the constraint on the initial
slice, the lapse and shift can be chosen {\em arbitrarily} on the
initial slice and thereafter.  These are referred to as gauge
choices\index{gauge conditions!in relativity}, analogous to the choice
of the gauge function $\Lambda$ in electrodynamics.  Einstein did not
specify these quantities; they are up to the numerical relativist to
choose at will.

\paragraph{Lapse.}
The choice of lapse corresponds to how one chooses 3D spacelike hypersurfaces
in the 4D spacetime.  The ``lapse'' of {\em proper} time along the normal
vector of one slice to the next is given by $\alpha \D t$, where $\D t$ is
the {\em coordinate} time interval between slices.  As $\alpha(x,y,z)$ can
be chosen at will on a given slice, some regions of spacetime can be made to
evolve farther into the future than others.

There are many motivations for particular choices of lapse.  A primary
concern is to ensure that it leads to a stable long term evolution.
It is easy to see that a naive choice of the lapse, e.g., $\alpha =1$,
the so-called geodesic slicing, suffers from a strong tendency to
produce coordinate singularities~\cite{Smarr78a,Smarr78b}.  A related
concern is
that one would like to cover the region of interest in an evolution,
say, where gravitational waves generated by some process could be
detected, while avoiding troublesome regions, say, inside black holes
where singularities lurk (the so-called ``singularity avoiding'' time
slicings).  Another important motivation is that some choices of
$\alpha$ allow one to write the evolution equations in forms that are
especially suited to numerical evolution.  Finally, computational
considerations also play important role in choice of the lapse; one
prefers a condition for $\alpha$ that does not involve great
computational expense, while also providing smooth, stable evolution.

Some ``traditional'' choices of the lapse used in the numerical
construction of spacetimes are~\cite{Piran83}: (1.) Lagrangian slicing,
in which the coordinates are following the flow of the matter in the
simulation.  This choice simplifies the matter evolution equations,
but it is not alway applicable, e.g., in a vacuum spacetime or when
the fluid flow pattern becomes complicated.  (2.) Maximal
slicing,~\cite{Smarr78a,Smarr78b} in which the trace of the extrinsic curvature
is required to be zero always, i.e, $K(t=0) = 0 = \partial_t K$.  The
evolution equations of the extrinsic curvature then lead to an
elliptic equation for the lapse
\begin{equation}
  D^a D_a \alpha - \alpha (R + K^2) =0 .
\label{maximal}
\end{equation}
The maximal slicing has the nice property of causing the lapse to
``collapse'' to a small value at regions of strong gravity, hence
avoiding the region that a curvature singularity is forming.  It is
one of the so-called ``singularity avoiding slicing conditions''.
Maximal slicing is easily the most studied slicing condition in
numerical relativity.  (3.)  Constant mean curvature, where we let $K
= constant$ different from zero, a choice often used in constructing
cosmological solutions.  (4.)  Algebraic slicing, where the lapse is
given as an algebraic function of the determinant of the three metric.
Algebraic slicing can also be singularity avoiding~\cite{Bona88}.  As
there is no need to solve an elliptic equation as in the case of
maximal slicing, algebraic slicing is computationally efficient.  Some
algebraic slicings (e.g., the harmonic slicing in which $\alpha$ is
set proportional to the square root of the determinant of the 3-metric
$g_{ab}$) also make the mathematical structure of the evolution
equations simpler.  However, the local nature of the choice of the
lapse could lead to noise in the lapse~\cite{Anninos94c} and the
formation of ``shock'' like features in numerical
evolutions~\cite{Alcubierre97a,Alcubierre97b}.  The former problem can
be dealt with by turning the algebraic slicing equation to an
evolution equation with a diffusion term~\cite{Anninos94c},
but the latter problem does not seem to have a simple solution.

In addition to these most widely used ``traditional'' choices of the lapse,
there are also some newly developed slicing conditions whose use in
numerical relativity though promising remain to be largely
unexplored~\cite{Tobias96} : (5.) K-driver.  This is a generalization of the
maximal slicing in which the extrinsic curvature, instead of being set
to zero, is required to satisfy the condition
\begin{equation}
  \partial_t K =  - c K ,
\label{kdrive}
\end{equation}
where c is some positive constant.  This was first brought up by
Eppley,~\cite{Eppley79} and recently investigated in
of the extrinsic curvature, when numerical inaccuracy causes it to
drift away from zero, is ``driven'' back to zero exponentially.  When
combined with the evolution equations, (\ref{kdrive}) again leads to
an elliptic equation for the lapse.  This choice of the lapse is
shown~\cite{Balakrishna96a} to lead to a much more stable numerical
evolution in cases where one wants to avoid large values of the
extrinsic curvature.  The optimal choice of the constant $c$ as well
as a number of variations on this ``driver'' scheme are presently being
studied.
(6.)  $\gamma$- driver.  This is another use of the ``driver'' idea.
In this case, the time rate of change of the determinant of the three
metric $det(g_{ab})$ is driven to zero ~\cite{Balakrishna96a}.  In the
absent of a shift vector or if the shift has zero divergence, this
reduces to the K-driver.  This choice of the lapse, which has the
unique property of being able to respond to the choice of the shift,
demands extensive investigations and evaluations.

\paragraph{Shift.}
The shift vector describes the ``shifting'' of the coordinates from
the normal vector as one moves from one slice to the next.  If the
shift vanishes, the coordinate point $(x,y,z)$ will move normal to a
given 3D time slice to the next slice in the future.  Please refer to
York~\cite{York79} or Cook~\cite{Cook90}, for details and diagrams.
The choice of shift is perhaps less well developed than the choice of
lapse in numerical relativity, and many choices need to be explored,
particularly in 3D. The main purpose of the shift is to ensure that
the coordinate description of the spacetime remains well behaved
throughout the evolution.  With an inappropriate or poorly chosen
shift, coordinate lines may move toward each other, or become very
stretched or sheared, leading to pathological behavior of the metric
functions that may be difficult to handle numerically.  It may even
cause the code to crash, if for example, two coordinate lines
``touch'' each other creating a ``coordinate singularity'' (i.e., the
metric becomes singular as the distance $ds$ between two coordinate
lines goes to zero).  Two important considerations for appropriate
shift conditions are the ability to prevent large shearing or drifting
of coordinates during an evolution, and the ability to control the
coordinate location of a physical object, e.g., the horizon of a black
hole.  These considerations are discussed below.  The development of
appropriate shift conditions for full 3D evolution, for systems
without symmetries, is an important research area that needs much
attention.  Geometrical shift conditions that can be formulated
without reference to specific coordinate systems or symmetries seem to
be desirable.  The basic idea is to develop a condition that minimizes
the stretching, shearing, and drifting of coordinates in a general
way.  A few examples have been devised which partially meet these
goals, such as ``minimal distortion'', ``minimal strain'', and
variations \cite{York79}, but much more investigations are needed.
New gauge conditions, based on these earlier proposals, have recently
been proposed but not yet tested in numerical
simulations~\cite{Brady98a}.

It is important to emphasize that the lapse and shift {\em only}
change the way in which the slices are chosen through a spacetime and
where coordinates are laid down on every slice, and do {\em not}, in principle,
affect any physical results whatsoever.  They {\em will} affect the
value of the metric quantities, but not the physics derived from them.
In this respect the freedom of choice in the lapse and shift is analogous
to the freedom of gauge in electromagnetic systems.

On the other hand, it is also important to emphasize that proper
choices of lapse and shift are crucial for the numerical construction
of a spacetime in the Einstein theory of general relativity, in
particular in a general 3D setting.  In a general 3D simulation
without symmetry assumption, there is no preferred choice of the form
of the metric (e.g., a diagonal 3-metric, or $g_{\theta \theta} = r ^
2 $ as in spherical symmetry), hence forcing us to deal with the gauge
degree of freedom in relativity in full.  This, when coupled with the
inevitable lower resolution in 3D simulations, often leads to
development of coordinate singularities, when evolved without a
sophisticated choice of lapse and shift.  Indeed the success of the
``driver'' idea suggested~\cite{Balakrishna96a} that in order to obtain a
stable
evolution over a long time scale, it is important to ensure that the
coordinate conditions used are not only suitable for the geometry of
the spacetime being evolved, but also that {\it the conditions
themselves are stable}. That is, when the condition is perturbed,
e.g., by numerical inaccuracy, there is no long term secular drifting.
We regard the construction of an algorithm for choosing a suitable
lapse and shift for a general 3D numerical simulation to be one of the
most important issues facing numerical relativity at present.

\subsection{General Relativistic Hydrodynamics}
In order to make numerical relativity a tool for computational general
relativistic astrophysics, it is important to combine numerical
relativity with traditional tools of computational astrophysics, and
in particular relativistic hydrodynamics.  While a large amount of 3D
studies in numerical relativity have been devoted to the {\em vacuum}
Einstein equations, the spacetime dynamics with a non-vanishing source
term remains a large uncharted territory.  As astrophysics of compact
objects that needs general relativity for its understanding is
attracting increasing attention, general relativistic hydrodynamics
will become an increasingly important subject as astrophysicists begin
to study more relativistic systems, as relativists become more
involved in studies of astrophysical sources.  This promises to be one
of the most exciting and important areas of research in relativistic
astrophysics in the coming years.

Previously, most work in relativistic hydrodynamics has been done on
fixed metric backgrounds.  In this approximation the fluid is allowed
to move in a relativistic manner in strong gravitational fields, say
around a black hole, but its effect on the spacetime is not
considered.  Over the last years very sophisticated methods for
general relativistic hydrodynamics have been developed by the Valencia
group led by Jos\'e M.
Ib\'a\~{n}ez~\cite{Marti91,Font94,Banyuls97,Donat98}.  These methods
are based on a hyperbolic formulation of the hydrodynamic equations,
and are shown to be superior to traditional artificial viscosity
methods for highly relativistic flows and strong shocks.

However, just fixed background approximation is inadequate in
describing a large class of problems which are of most interest to
gravitational wave astronomy, namely those with substantial matter
motion generating gravitational radiation, like the coalescences of
neutron star binaries.  As will be discussed in more detail below, we
are constructing a multi-purpose 3D code for the NASA Neutron Star
Grand Challenge Project~\cite{Nasa98} that contains the full Einstein
equations coupled to general relativistic hydrodynamics.  The
spacetime part of the code is based on the ``Cactus'' code; the
hydrodynamic part consists of both an artificial viscosity
module,~\cite{Wang97} and a module (MAHC HYPERBOLIC\_HYDRO) based on
modern shock capturing schemes~\cite{Font98b}.

The ``MAHC'' general relativistic hydro code at present contains three
hydro evolution methods~\cite{Font98b}: a flux split method, Roe's
approximate Riemann solver~\cite{Roe81} and Marquina's approximate
Riemann solver~\cite{Donat96,Donat98}.  All three are based on
finite-difference schemes employing approximate Riemann solvers to
account explicitly for the characteristic information of the
equations.  These schemes are particularly suitable for astrophysics
simulations that involve matter in (ultra)relativistic speeds and
strong shock waves.

In the flux split method, the flux is decomposed into the part
contributing to the eigenfields with positive eigenvalues (fields
moving to the right) and the part with negative eigenvalues (fields
moving to the left).  These fluxes are then discretized with one sided
derivatives (which side depends on the sign of the eigenvalue).  The
flux split method presupposes that the equation of state of the fluid
has the form $P = P(\rho,\epsilon) = \rho f(\epsilon)$, which
includes, e.g., the adiabatic equation of state.  The second scheme,
Roe's approximate Riemann solver~\cite{Roe81} is by now a
``traditional'' method for the integration of non-linear hyperbolic
systems of conservation laws.~\cite{Font94,Eulderink94,Banyuls97} This
method makes no assumption on the equation of state, and, is more
flexible than the flux split methods.  The third method, the
Marquina's Method, is a promising new scheme.\cite{Donat96} It is
based on a flux formula which is an extension of Shu and Osher's
entropy-satisfying numerical flux~\cite{Shu89} to systems of
hyperbolic conservation laws.  In this scheme there are no artificial
intermediate states constructed at each cell interface.  This implies
that there are no Riemann solutions involved (either exact or
approximate); moreover, the scheme has been proved to alleviate
several numerical pathologies associated to the introduction of an
{\it averaged} state (as Roe's method does) in the local
diagonalization procedure (see~\cite{Donat96,Donat98}).  For a
detailed comparison of the three schemes and their coupling to
dynamical evolution of spacetimes, see ~\cite{Font98b}.

The availability of the hyperbolic hydro treatment and its coupling to
the spacetime evolution code is particularly noteworthy.  With the
development of a hyperbolic formulation of the Einstein equations
described above, the {\em entire} system can be treated as a single
system of hyperbolic equations, rather than artificially separating
the spacetime part from the fluid part.  Such a unified treatment
based on the ``MAHC'' module is presently under construction by our
NCSA/Potsdam/WashU collaboration.


\subsection{Boundary Conditions}
Appropriate conditions for the outer boundary have yet to be derived
for 3D numerical relativity. In 1D and 2D relativity codes, the outer
boundary is generally placed far enough away that the spacetime is
nearly flat there, and static or flat (i.e., copying data from the
next-to-last zone to the outer edge) boundary conditions can usually
be specified for the evolved functions.  However, due to the
constraints placed on us by limited computer memory, this is not
currently possible in 3D. Adaptive mesh refinement will be of great
use in this regard, but will not substitute for proper physical
treatment.  Most results to date have been computed with the evolved
functions kept static at the outer boundary, even if the boundaries
are too close for comfort in 3D!

There are several other approaches under development that promise to
improve this situation greatly that we will not have room to explore in
detail here, but should be mentioned.  Generally, one has in mind
using Cauchy evolution in the strong field, interior region where, say,
black holes are colliding.  This outer part of this region will be
matched to some exterior treatment designed to handle what is
primarily expected to be outgoing radiation.

Two major approaches have been developed by the NSF Black Hole Grand
Challenge Alliance, a large US collaboration working to solve the
black hole coalescence problem, and other groups.  First, by using
perturbation theory, it is possible to identify quantities in the
numerically evolved metric functions that obey the Regge-Wheeler and
Zerilli wave equations that describe gravitational waves propagating
on a black hole background.  These can be used to provide boundary
conditions on the metric and extrinsic curvature functions in an
actual evolution, as described in a recent paper~\cite{Abrahams97a}.
This is an excellent step forward in outer boundary treatments that
should work to minimize reflections of the outgoing wave signals from
the outer boundary.  In tests with weak waves, a full 3D Cauchy
evolution code has been successfully matched to the perturbative
treatment at the boundary, permitting waves to escape from the
interior region with very little reflection.  Alternatively,
``Cauchy-Characteristic matching'' attempts to match spacelike slices
in the Cauchy region to null slices at some finite radius, and the
null slices can be carried out to null infinity.  3D characteristic
evolution codes have progressed dramatically in recent years, and
although the full 3D matching remains to be completed, tests of the
scheme in specialized settings show promise\cite{Bishop98a}.  One can
also use the hyperbolic formulations of the Einstein equations to find
eigenfields, for which outgoing conditions can in principle be
applied\cite{Bona94b} in 1D.  In 3D this technique is still under
development, but it shows promise for future work.  Finally, there is
another hyperbolic approach which uses conformal rescaling to move the
boundary to
infinity~\cite{Friedrich81a,Friedrich81b,Friedrich96,Huebner96}.
These methods have different strengths and weaknesses, but all promise
to improve boundary treatments significantly, helping to enable longer
evolutions than are presently possible.

\subsection{Special Difficulties with Black Holes}

The techniques described so far are generic in their application in
numerical relativity.  But in this section we describe a few problems
that are characteristic of black holes, and special algorithms under
development to handle them.  Black hole spacetimes all have in common
one problem: singularities lurk within them, which must be handled
numerically.  Developing suitable techniques for doing so is one of
the major research priorities of the community at present.  If one
attempts to evolve directly into the singularity, infinite curvature
will be encountered, causing any numerical code to break down.

\begin{figure}
\label{singularity}
\hspace{1.5cm}\epsfysize=8cm
\epsfbox{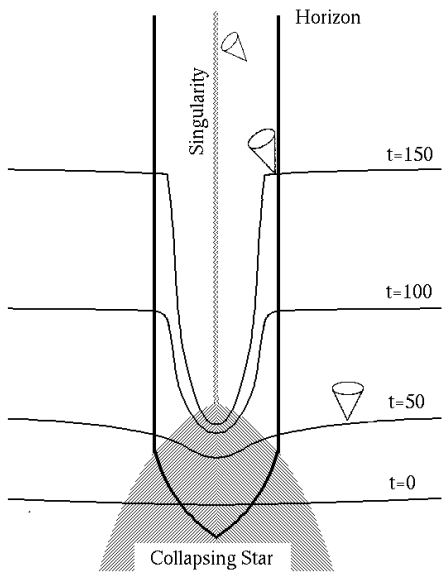}
\caption{A spacetime
diagram showing the formation of a BH, and time slices
traditionally used to foliate the spacetime in traditional numerical
relativity with singularity avoiding time slices.  As the evolution
proceeds, pathologically warped hypersurfaces develop, leading to
unresolvable gradients that cause numerical codes to crash.}
\end{figure}

Traditionally, the singularity region is avoided by the use of
``singularity avoiding'' time slices, that wrap up around the
singularity.  Consider the evolution shown in Fig.~\ref{singularity}.
A star is collapsing, a singularity is forming, and time slices are
shown which avoid the interior while still covering a large fraction
of the spacetime where waves will be seen by a distant observer.
However, these slicing conditions by themselves do not solve the
problem; they merely serve to delay the onset of instabilities.  As
shown in Fig.~\ref{singularity}, in the vicinity of the singularity
these slicings inevitably contain a region of abrupt change near the
horizon, and a region in which the constant time slices dip back deep
into the past in some sense.  This behavior typically manifests itself
in the form of sharply peaked profiles in the spatial metric
functions~\cite{Smarr78b}, ``grid stretching''~\cite{Shapiro86} or large
coordinate shift~\cite{Bernstein89} on the BH throat, etc.
Numerical simulations will eventually fail due to these pathological
properties of the slicing.

\subsubsection{Apparent Horizon Boundary Conditions (AHBC)}
The cosmic censorship conjecture suggests that in physical situations,
singularities are hidden inside BH horizons.  Because the region of
spacetime inside the horizon is causally disconnected from the region
of interest outside the horizon, one is tempted numerically to cut
away the interior region containing the singularity, and evolve only
the singularity-free region outside, as originally suggested by
Unruh\cite{Unruh84}.  This has the consequence that there will be a
region inside the horizon that simply has no numerical data.  To an
outside observer no information will be lost since the regions cut
away are unobservable.  Because the time slices will not need such
sharp bends to the past, this procedure will drastically reduce the
dynamic range, making it easier to maintain accuracy and stability.
Since the singularity is removed from the numerical spacetime, there
is in principle no physical reason why BH codes cannot be made to run
indefinitely without crashing.

We spoke innocently about the BH horizon, but did not distinguish
between the {\em apparent} and {\em event} horizon.  These are very
different concepts!  While the event horizon, which is roughly a null
surface that never reaches infinity and never hits the singularity,
may hide singularities from the outside world in many situations,
there is no guarantee that the apparent horizon, which is the
(outermost) surface that has instantaneously zero expansion
everywhere, even exists on a given slice!  (By ``zero expansion'' we
mean that the surface area of outgoing bundles of photons normal to
the surface is constant.  Hence, the surface is ``trapped.'')  Methods
for finding event horizons in numerical spacetimes are now known, and
will be discussed below.  But event horizons can only be found after
examining the {\em history} of an evolution that has been already been
carried out to sufficiently late times\cite{Anninos94f,Libson94a}.
Hence they are useless in providing boundaries as one integrates {\em
forward} in time.  On the other hand the apparent horizon, if it
exists, can be found on any given slice by searching for closed
2--surfaces with zero expansion.  Although one should worry that in a
generic BH collision, one may evolve into situations where no apparent
horizon actually exists, let us cross that bridge if we come to it.
Methods for finding apparent horizons will also be discussed below,
but for now we assume that such a method exists.

Given these considerations, there are two basic ideas behind the
implementation of the apparent horizon boundary condition (AHBC),
also known as black hole excision:

{\em (a)} It is important to use a finite differencing scheme which
respects the causal structure of the spacetime.  Since the horizon is
a one-way membrane, quantities on the horizon can be affected only by
quantities outside but not inside the horizon: all quantities on the
horizon can in principle be updated solely in terms of known
quantities residing on or outside the horizon.  
There are various
technical details and variations on this idea, which is called
``Causal Differencing''\cite{Seidel92a} or ``Causal
Reconnection''\cite{Alcubierre94a}, but here we focus primarily on the
basic ideas and results obtained to date.  

{\em (b)} A shift is used to control the motion of the horizon, and
the behavior of the grid points outside the BH, as they tend to fall
into the horizon if uncontrolled.

An additional advantage to using causal differencing is that it allows
one to follow the information flow to create grid points with proper
data on them, as needed inside the horizon, even if they did not exist
previously.  (Remember above that we have cut away a region inside the
horizon, so in fact we have no data there.)  One example is to let a
BH move across the computational grid.  If a BH is moving physically,
it may also be desirable to have it move through coordinate space.
Otherwise, all physical movement will be represented by the ``motion''
of the grid points.  For a single BH moving in a straight line, this
may be possible (though complicated), but for spiraling coalescence
this will lead to hopelessly contorted grids.  The immediate
consequence of this is that as a BH moves across the grid, regions in
the wake of the hole, now in its exterior, must have previously been
inside it where no data exist!  But with AHBC and causal differencing
this need not be a problem.

Does the AHBC idea work?  Preliminary indications are very promising.
In spherical symmetry (1D), numerous studies show that one can locate
horizons, cut away the interior, and evolve for essentially unlimited
times ($t \propto 10^{3-4}M$, where $M$ is the black hole mass).  The
growth of metric functions can be completely controlled, errors are
reduced to a very low level, and the results can be obtained with a
large variety of shift and slicing conditions, and with matter falling
in the BH to allow for true dynamics even in spherical
symmetry\cite{Seidel92a,Anninos94e,Scheel94,Marsa96}.

In 3D, the basic ideas are similar but the implementation is much more
difficult.  The first successful test of these ideas to a
Schwarzschild BH in 3D used horizon excision and a shift provided from
similar simulations carried out with a 1D code\cite{Anninos94c}.  The
errors were found to be greatly reduced when compared even to the 1D
evolution with singularity avoiding slicings.  Another 3D
implementation of the basic technique was provided by
Br{\"u}gmann~\cite{Bruegmann96}.

This was a proof of principle, but more general treatments are
following.  Daues extended this work to a full range of shift
conditions~\cite{Daues96a}, including the full 3D minimal distortion
shift~\cite{York79}.  He also applied these techniques to dynamic
BH's, and collapse of a star to form a BH, at which point the horizon
is detected, the region interior to the horizon excised, and the
evolution continued with AHBC. The focus of this work has been on
developing general gauge conditions for single BH's without movement
through a grid.  Under these conditions, BH's have been accurately
evolved well beyond $t=100M$.  The NSF Black Hole Grand Challenge
Alliance had focussed on development of 3D AHBC techniques for
moving Schwarzschild BH's\cite{Cook97a}.  In this work, analytic gauge
conditions are provided, which are chosen to make the evolution
static, although the numerical evolution is allowed to proceed freely.
This moving hole is the first successful 3D test of populating grid
points with data as they emerge in the BH wake.

These new results are significant achievements, and show that the
basic techniques outlined above are not only sound, but are also
practically realizable in a 3D numerical code.  However, there is
still a significant amount of work to be done!  The techniques have
yet to be applied carefully to distorted BH's, with tests of
the waveforms emitted.  They have not been applied to
rotating BH's of any kind; they have not been applied to
colliding BH's with horizon topology change, and moving black
holes have yet to be evolved in AHBC with a nonanalytic gauge choice.
There are still clearly many steps to be taken before the techniques
will be successfully applied to the general BH merger problem.

\section{Tools for Analyzing the Numerical Spacetimes}
We now turn to the description of several important tools that have
been developed to analyze the results of a numerical evolution,
carried out by some numerical evolution scheme.  The evolution will
generally provide metric functions on a grid, but as described above
these functions are highly dependent on both the coordinate system and
gauge in which the system is evolved.  Determining {\em physical}
information, such as whether a black hole exists in the data, or what
gravitational waveforms have been emitted, are the subjects of this
section.

\subsection{Horizon Finders}

As described above, black holes are defined by the existence of an
event horizon (EH), the surface of no return from which nothing, not
even light, can escape.  The event horizon is the boundary that
separates those null geodesics that reach infinity from those that do
not.  The global character of such a definition implies that the
position of an EH can only be found if the whole history of the
spacetime is known.  For numerical simulations of black hole
spacetimes in particular, this implies that in order to locate an EH
one needs to evolve sufficiently far into the future, up to a time
where the spacetime has basically settled down to a stationary
solution.  Recently, methods have been developed to locate and analyze
black hole horizons in numerically generated spacetimes, with a number
of interesting results
obtained~\cite{Anninos94f,Libson94a,Hughes94a,Matzner95a,Masso95a,Shapiro95a}.

In contrast, an apparent horizon (AH) is defined locally in time as the
outer-most marginally trapped surface~\cite{Hawking73a}, i.e. a surface
on which the expansion of out-going null geodesics is everywhere zero.  An AH
can therefore be defined on a given spatial hypersurface.  A well known
result~\cite{Hawking73a} guarantees that if an AH is found, an EH must
exist somewhere outside of it and hence a black hole has formed.

\subsection{Locating the apparent horizons}

The expansion $\Theta$ of a congruence of null rays moving in the
outward normal direction to a closed surface can be shown to
be~\cite{York89}
\begin{equation}
\Theta = \nabla_i s^i + K_{ij} s^i s^j - {\rm tr} K ,
\label{eqn:expansion}
\end{equation}
where $\nabla_i$ is the covariant derivative associated with the
3-metric $\gamma_{ij}$, $s^{i}$ is the normal vector to the surface,
$K_{ij}$ is the extrinsic curvature of the time slice, and ${\rm tr}
K$ is its trace. An AH is then the outermost surface such that
\begin{equation}
\Theta = 0.
\label{eqn:horizon1}
\end{equation}
This equation is not affected by the presence of matter, since it is
purely geometric in nature.  We can use the same horizon
finders without modification for vacuum as well as non-vacuum spacetimes.
The key is to find a closed surface with normal vector $s^{i}$ satisfying
this equation.

\subsubsection{Minimization Algorithms}
As apparent horizons are defined by the vanishing of the expansion on
a surface, a fairly obvious algorithm to find such a surface involves
minimizing a suitable norm of the expansion below some tolerance while
adjusting test surfaces.  Minimization algorithms for finding apparent
horizons were among the first methods
developed~\cite{Brill63,Eppley77}.  More recently, a
3D minimization algorithm was developed and implemented by the
Potsdam/NCSA/WashU group, applied to a variety of black hole initial
data and 3D numerically evolved black hole
spacetimes~\cite{Libson94b,Libson95a,Libson93a,Camarda97a,Camarda97c}.
Essentially the same algorithm was also implemented independently by
Baumgarte {\em et.al.}~\cite{Baumgarte96}.

The basic idea behind a minimization algorithm is to assume the
surface can be represented by a function $F(x^i)=0$, expand it the in
terms of some set of basis functions, and then minimize the integral
of the square of the expansion $\Theta^2$ over the surface.  For
example, one can parameterize a surface as
\begin{equation}
F(r,\theta,\phi) = r - h(\theta,\phi) .
\label{eqn:F}
\end{equation}
The surface under consideration will be taken to correspond to the
zero level of $F$.  The function $h(\theta,\phi)$ is then expanded
in terms of spherical harmonics:
\begin{equation}
h(\theta,\phi) = \sum_{l=0}^{l_{\rm max}} \sum_{m=-l}^{l}
a_{lm} Y_{lm}(\theta,\phi) .
\end{equation}
Similar techniques were developed by~\cite{Nakamura84}.

At an AH the expansion integral over the surface should
vanish, and we will have a global minimum.  Of course, since
numerically we will never find a surface for which the integral
vanishes exactly, one must set a given tolerance level below which a
horizon is assumed to have been found.

Minimization algorithms for finding AH's have a few drawbacks: First,
the algorithm can easily settle down on a local minimum for which the
expansion is not zero, so a good initial guess is often required.
Moreover, when more than one marginally trapped surface is present, as
is often the case, it is very difficult to predict which of these
surfaces will be found by the algorithm: The algorithm can often
settle on an inner horizon instead of the true AH. Again, a good
initial guess can help point the finder towards the correct horizon.
Finally, minimization algorithms tend to be very slow when compared
with `flow' algorithms of the type described in the next section.
Typically, if $N$ is the total number of terms in the spectral
decomposition, a minimization algorithm requires of the order of a few
times $N^2$ evaluations of the surface integrals (where in our
experience `a few' can sometimes be as high as 10).

This algorithm has been implemented in the ``Cactus'' code for 3D
numerical relativity~\cite{Bona98b}.  For more details of the
application of this algorithm, see
Refs.~\cite{Libson94b,Libson95a,Baumgarte96,Libson93a}.

\subsubsection{3D fast flow algorithm}

A second method that has been implemented in the ``Cactus'' code is the
``fast flow'' method proposed by Gundlach~\cite{Gundlach97a}. Starting
from an initial guess for the $a_{lm}$, it approaches the
apparent horizon through the iteration
\begin{equation}
\label{SpectralAHF}
a_{lm}^{(n+1)}
= a_{lm}^{(n)}
-{A\over 1 + B l(l+1)}
\left(\rho \Theta \right)_{lm}^{(n)}.
\end{equation}
where $(n)$ labels the iteration step, $\rho$ is some positive
definite function (``a weight''), and $(\rho \Theta)_{lm}$ are the harmonic
components of the function $(\rho \Theta)$. Various choices for the weight
$\rho$ and the coefficients $A$ and $B$ parameterize a family of such
methods. The fast flow algorithm in Cactus uses
\begin{equation}
\label{rho}
\rho = 2 \; r^2 |\nabla  F| \left[ \left(g^{ij}-s^i s^j \right)
\left( \bar g_{ij}-\nabla_i r \nabla_j r \right) \right]^{-1} ,
\end{equation}
where $\bar g_{ij}$ is the flat background metric associated with the
coordinates $(r,\theta,\phi)$, and
\begin{equation}
\label{alphabeta}
A = {\alpha\over l_{\rm{max}}(l_{\rm{max}}+1)} + \beta,
\quad
B = {\beta\over \alpha}.
\end{equation}
with $\alpha=c$ and $\beta=c/2$. Here $c$ is a variable step size,
with a typical value of $c\sim 1$. $l_{\rm{max}}$ is the maximum
value of $l$ one chooses to use in describing the surface.  The
iteration procedure is a finite difference approximation to a
parabolic flow, and the adaptive step size is chosen to keep the
finite difference approximation roughly close to the flow limit to
prevent overshooting of the true apparent horizon. The adaptive step
size is determined by a standard method used in ODE integrators: we
take one full step and two half steps and compare the resulting
$a_{lm}$. If the two results differ too much one from another, the
step size is reduced.

Other methods for finding apparent horizons, based directly on
computing the jacobian of the finite differenced horizon equation,
have been developed\cite{Thornburg95,Huq95} and successfully used in
3D. For details, please see these references.

\subsection{Locating the event horizons}

The AH is defined locally in time and hence is much easier to locate
than the event horizon (EH) in numerical relativity.  The EH is a
global object in time; it is traced out by the path of outgoing light
rays that {\em never} propagate to future null infinity, and {\em
never} hit the singularity.  (It is the boundary of the causal past of
future null infinity $\dot{\cal{J}}^{{-}}(\cal{I}^{+})$.)  In
principle one needs to know the entire time evolution of a spacetime
in order to know the precise location of the EH. However, in spite of
the global properties of the EH, hope is not lost for finding it very
accurately, even in a numerical simulation of finite duration.  Here
we discuss a method to find the EH, given a numerically constructed
black hole spacetime that eventually settles down to an approximately
stationary state at late times.  In principle, as the EH is a null
surface, it can be found by tracing the path of null rays through
time.  Outward going light rays emitted just outside the EH will
diverge away from it, escaping to infinity, and those emitted just
inside the EH will fall away from it, towards the singularity.  In a
numerical integration it is difficult to follow accurately the
evolution of a horizon generator forward in time, as small numerical
errors cause the ray to drift and diverge rapidly from the true EH. It
is a physically unstable process.  But we can actually use this
property to our advantage by considering the time-reversed problem.
In a global sense in time, any outward going photon that begins near
the EH will be {\em attracted} to the horizon if integrated {\em
backward} in time~\cite{Anninos94f,Libson94b}.  In integrating
backwards in time, it turns out that it suffices to start the photons
within a fairly broad region where the EH is expected to reside.  Such
a horizon-containing region, as we call it, is often easy to determine
after the spacetime has settled down to approximate stationarity.  The
crucial point is that when integrated backward in time along null
geodesics, this horizon-containing region shrinks rapidly in
``thickness'', leading to a very accurate determination of the
location of the EH at earlier times.  Note that it is the earlier time
when the black hole is under highly dynamical evolution that we are
really interested in.

Although one can integrate individual null geodesics backward in time, we
find that there are various advantages to integrate the entire null
surface backward in time.  A null surface, if defined by
$f(t,x^i)=0$ satisfies the condition
\begin{equation}
g^{\mu\nu} \partial_{\mu} f \partial_{\nu} f = 0\; .
\label{nullsurface}
\end{equation}
Hence the evolution of the surface can be obtained by a simple
integration,
\begin{equation}
\partial_t f = \frac{ - g^{ti} \partial_i f +
\sqrt{(g^{ti}\partial_i f)^2 - g^{tt} g^{ij} \partial_i f \partial_j f}
}{g^{tt}} \;  .
\label{evolveh}
\end{equation}
The inner and outer boundary of the horizon containing region when
integrated backward in time, will rapidly converge to practically a
single surface to within the resolution of the numerically constructed
spacetime, i.e., a small fraction of a grid point.  An accurate
location of the event horizon is hence obtained.  We henceforth shall
represent the horizon surface as the function $f _ H (t,x^i)$.  Aside
from the simplicity of this method, there are a number of
technical advantages as discussed in~\cite{Anninos94f}.  One
particularly noteworthy point is that this method is capable of
giving the caustic structure of the event horizon if there is any; for
details see~\cite{Anninos94f}.

The function $f _ H (t,x^i)$ provides the complete coordinate location
of the EH through the spacetime (or a very good approximation of it,
as shown in~\cite{Libson94a}).  This function by itself directly
gives us the topology and location of the EH. When combined with the
induced metric function on the surface, which is recorded throughout
the evolution, it gives the intrinsic geometry of the EH. When further
combined with the spacetime metric, all properties of the EH including
its embedding can be obtained\index{horizons!event}.  Moreover, as the
normal of $f _ H (t,x^i) =0 $ gives the null generators of the
horizon, it is an easy further step to determine the null generators,
and hence the complete dynamics of the horizon in this formulation.

As described in a series of papers, the event horizon, once found with
such a method, can be analyzed to provide important information about
the dynamics of black holes in a numerically generated
spacetime~\cite{Anninos94f,Libson94a,Hughes94a,Matzner95a,Masso95a,Shapiro95a}.

\subsection{Wave Extraction}
The gravitational radiation emitted is one of the most important
quantities of interest in many astrophysical processes.  The radiation
is generated in regions of strong and dynamic gravitational fields,
and then propagated to regions far away where it will someday be
detected.  We take the approach of computing the generation and
evolution of the fields in a fully nonlinear way, while analyzing the
radiation with a perturbation formulation in the regions where it can be
so treated.

The theory of black hole perturbations is well developed.  One
identifies certain perturbed metric quantities that evolve according
to wave equations on the black hole background.  These
perturbed metric functions are also dependent of the gauge in which
they are computed.  We use a {\em gauge-invariant}
prescription for isolating wave modes on black hole background,
developed first by Moncrief~\cite{Moncrief74}.  The basic idea is that
although the perturbed metric functions transform under
coordinate transformations (gauge transformations), one
can identify certain linear combinations of these functions that are
invariant to first order of the perturbation.  These gauge-invariant functions
are the quantities that carry true physics, which does not
depend on coordinate systems. They obey the
wave equations describing waves propagating on the fixed blackhole background.
There are two independent wave modes, even- and odd-parity,
corresponding to the two degrees of freedom, or polarization modes, of
the waves.

A {\em waveform extraction} procedure has been developed that allows one to
process the metric and to identify the wave modes.  The gravitational
wave function (often called the ``Zerilli function'' for even-parity
or the ``Regge-Wheeler function'' for odd-parity) can be computed by
writing the metric as the sum of a background black hole part and a
perturbation:
\begin{equation}
g_{\alpha\beta}=\stackrel{o}{g}_{\alpha\beta}+h_{\alpha
\beta}(Y_{\ell,m}),
\end{equation}
where the perturbation $h_{\alpha\beta}$ is expanded in spherical
harmonics and their tensor generalizations and the background part
$\stackrel{o}{g}_{\alpha\beta}$ is spherically symmetric.  To compute
the elements of $h_{\alpha\beta}$ in a numerical simulation, one
integrates the numerically evolved metric components $g_{\alpha\beta}$
against appropriate spherical harmonics over a coordinate 2--sphere
surrounding the black hole, making use of the orthogonality properties
of the tensor harmonics.  This process is performed for each $\ell ,
m$ mode for which waveforms are desired.  The resulting functions
$h_{\alpha \beta}(Y_{\ell,m})$ can then be combined in a
gauge-invariant way, following the prescription given by
Moncrief\cite{Moncrief74}.  For each $\ell , m$ mode, this gauge
invariant gravitational waveform can be extracted when the wave passes
through ``detectors'' at some fixed radius in the computational grid.
This procedure has been described in detail
in~\cite{Brandt94a,Brandt94b,Brandt94c}, and more generally in
Refs.~\cite{Allen97a,Allen98a,Camarda97c}.  It works amazingly well,
allowing extraction of waves that carry very small energies (of order
$10^{-7}M$ or less, with $M$ being the mass of the source) away from
the source.  The procedure should apply to any isolated source of
waves, such as colliding black holes, neutron stars, etc.  If the
sources are rotating, this procedure should be generalized to use the
Teukolsky formalism describing perturbations about a Kerr black hole,
but this has not yet been done.  Instead, the spherical perturbation
theory (with a few minor modifications) has been applied to distorted
rotating black holes with satisfactory
results~\cite{Brandt94a,Brandt94b,Brandt94c}.

\section{Computational Science, Numerical Relativity, and the ``Cactus''
Code}

\subsection{The Computational Challenges of Numerical Relativity}

Before we describe our computational methods in the following
subsections, we summarize the computational challenges of numerical
relativity discussed above.  It is in response to these challenges
that we have devised the computational methods.

\noindent $\bullet$ Computational challenges due to the complexity of
the physics involved: The Einstein equations are probably the most
complex partial differential equations in all of physics, forming a
system of dozens of coupled, nonlinear equations, with thousands of
terms, of mixed hyperbolic, elliptic, and even undefined types in a
general coordinate system.  The evolution has elliptic constraints
that should be satisfied at all times.  In simulations without
symmetry, as would be the case for realistic astrophysical processes,
codes can involve hundreds of 3D arrays, and ten of thousands of
operations per grid point per update.  Moreover, for simulations of
astrophysical processes, we will ultimately need to integrate
numerical relativity with traditional tools of computational
astrophysics, including hydrodynamics, nuclear astrophysics, radiation
transport and magneto-hydrodynamics, which govern the evolution of the
source terms (i.e., the right hand side) of the Einstein equations.
This complexity requires us to push the frontiers of massively
parallel computation.

\noindent $\bullet$ Challenge in Collaborative Technology: The
integration of numerical relativity into computational astrophysics is
a multi-disciplinary development, partly due to the complexity of the
Einstein equations, and partly due to the physical systems of
interest.  Solving the Einstein equations on massively parallel
computers involves gravitational physics, computational science,
numerical algorithm and applied mathematics.  Furthermore, for the
numerical simulations of realistic astrophysical systems, many physics
disciplines, including relativity, astrophysics, nuclear physics, and
hydrodynamics are involved.  It is therefore essential to have the
numerical code software engineered to allow co-development by
different research groups {\it and} groups with different expertise.

\noindent $\bullet$ The numerical construction of a spacetime itself
presents unique challenges: According to the singularity theorems of
general relativity, regions of strong gravity often generate spacetime
singularities.  Due to the need to avoid spacetime
singularities~\cite{Seidel96a,Seidel96b}, and to obtain long term
stability in the numerical simulations, sophisticated control of the
coordinate system is needed for the construction of a numerical
spacetime.  This dynamic interplay between the spacetime being
constructed and the computational coordinate choice itself (``gauge
choice'') is a unique feature of general relativity that makes the
numerical simulations much more demanding.  Besides extra
computational power, advanced visualization tools, preferably real
time interactive ``window into the oven'' visualization, are
particularly useful in the numerical construction.

\noindent $\bullet$ The multi-scale problem: Astrophysics of strongly
gravitating systems inherently involves many length and time scales.
The microphysics of the shortest scale (the nuclear force), controls
macroscopic dynamics on the stellar scale, such as the formation and
collapse of neutron stars (NSs).  On the other hand, the stellar scale
is at least 10 times {\it less} than the wavelength of the
gravitational waves emitted, and many orders of magnitude less than
the astronomical scales of their accretion disk and jets; these larger
scales provide the directly observed signals.  Numerical studies of
these systems, aiming at direct comparison with observations,
fundamentally require the capability of handling a wide range of
dynamical time and length scales.

All of these issues lead to important research questions in
computational science.  Here we give an overview of some of our effort
in these directions, focusing on performance and coding issues on
parallel machines, and on the development of a community code that
incorporates all the mathematical and computational techniques
described above (and many more), in a collaborative infrastructure for
numerical relativity.

\subsection{Code Generation and Data Parallel Fortran}
When expanded out in a particular coordinate system the evolution
equations for the full Einstein equations in the 3$+$1 formulation
have many thousands of terms.  These are usually derived and coded in
Fortran with a symbolic manipulator package such as Mathematica or
Macsyma.  However, these packages often generate Fortran expressions
that are unsuitable for most compilers, even on traditional
supercomputers.  We often exceed internal compiler limits on length of
expression, number of continuation lines, number of arguments to a
subroutine, number of nested parentheses, and so forth.  So our code
generation techniques need to be carefully massaged before an
efficient, working code is generated.

The evolution equations are generally written out using explicit 
finite difference schemes, which are very popular for hyperbolic 
systems of equations.  These equations are good candidates for the 
``SIMD'' style approach to programming parallel machines.  (SIMD 
stands for ``Single Instruction Multiple Data'', which means an 
operation like ``add arrays A and B together'' can be carried out 
completely in parallel, with the same instruction (add) on multiple 
data elements in memory.  This is also a so-called ``data parallel'' 
operation, since the same operation is applied simultaneously to all 
data elements of arrays A and B in parallel, and no communications are 
required between processors.)  Until recently, in our research group 
3D codes have been generally written in this style using data parallel 
Fortran90 and CMFortran style languages.  With this approach, 
communications between processors, required for example when computing 
derivatives (which require knowledge of neighboring data points in 
memory), are handled by the compilers without need for the user to do 
anything.  We have used the C-preprocessor to incorporate a few 
different code blocks so that we can maintain a single source file for 
several machines.  (For an excellent review of many modern approaches 
to parallel computing, including further information on many of the 
concepts and acronyms common in computational science, see, 
e.g.,~\cite{Foster95a}, available both in print and on-line).

Using this global approach we previously developed a single code,
called H3expresso, that achieved over 15 Gflops on the 512 node CM-5
and about 8 Gflops on the 16 processor Cray C-90.  This code was one
of the fastest applications on either machine~\cite{Hillis93}.  We
performed a detailed comparative study of this code on many
architectures, including the C-90, Convex SPP-1200, T3D, CM-5, SGI
Power Challenge, and SP-2, achieving excellent scaling all machines.
These results are possible because of the very high
computation/communication ratio inherent in the Einstein equations.
The hyperbolic equations contain thousands of terms to be evaluated,
while the only communications required are in computing finite
differences for numerical derivatives.

\subsection{Data Parallel Fortran Evolves to MPI}
However, this data parallel approach is not the best one to follow 
with more modern microprocessor based scalable supercomputers, such as 
the SGI/Cray Origin 2000 and Cray T3D and T3E, due largely to the use 
of caches that boost performance of a single node.  It is worth 
commenting on how we have adapted the H3expresso code to a ``message 
passing'' language like MPI, with single processor optimizations, 
which then led to to the development of the new Cactus code described 
below.  (MPI stands for ``Message Passing Interface'', a standard 
communications library now available on most parallel computers, that 
allows the user to explicitly control the communication of data 
between processors when required~\cite{Foster95a}.  This can be more 
efficient than allowing the compiler to handle this automatically.)

Due to the data-parallel nature of the code, many of the temporaries
evolved in solving the hyperbolic equations (\ref{3Dbalance}), notably
the sources and the fluxes, are created as 3D arrays. This allows
fairly easy parallelization of the code with MPI. Since the only
finite differencing in the code is on the fluxes, they are the only
variables which need communication, and thus we can easily do an
MPI-based communication with these variables during our update loop.

Unfortunately, one of the major problems of the data parallel
programming model is that it requires the creation of large numbers of
3D temporary arrays to store source or flux terms.  On a system like
the CM-5, this technique was crucial in obtaining performance; the
arrays were distributed and were stored on the vector units, so the
system could operate on them quickly and communicate them
transparently.  However, with single statements for entire arrays with
large degrees of complexity, each assignment requires a sweep through
the complete memory space.  Cache locality is impossible, and the code
performs very poorly in an out-of-cache regime.  Hence, to achieve
high single processor performance on more modern microprocessor based
architectures special attention has to be paid to rewriting
expressions to maximize the use of the processor cache.

Using the experience gained from exploring issues of parallelism and 
single processor performance with the H3expresso code, we have 
developed from the scratch a new 3D Einstein equation code, the 
``Cactus'' code, which integrates important design decisions for 
modern HPC (standard acronym for ``High Performance Computing'') 
architectures from the outset:
\begin{itemize}
\item The numerical kernals for the Einstein equations, needed by all
users, are highly optimized for modern microprocessor based
architecture.
\item Other routines (e.g., waveform extraction) are
written by the community of users in either C or Fortran.
\item It obtains parallelism through MPI, not through compiler technologies.
In the following, we describe some details of single processor
performance, parallelism and the code's collaborative
infrastructure.
\end{itemize}

\subsubsection{Parallelism using MPI}
In ``Cactus'', we achieve parallelism using ghost-zone domain
decomposition. That is, we decompose a global domain over our
processors, and place an overlap region on each processor. Then each
single processor is responsible for updating the interior of their
local region, and a MPI communication synchronizes the boundary zones
after communications. We have also optimized for the particular
structure of our equations. The straightforward way to set up our
communication patterns would be an algorithm like
\begin{verbatim}
  for (it = 0 to maxit) {
    update interior for a time step dt
    update ghost zones for all grid functions
  }
\end{verbatim}
However, many of our variables have no flux.  Since we generally use a
strang split in the hyperbolic evolution, which allows us to update
our source and flux evolution separately, the integration of these
flux-free variables is a completely spatially de-coupled point local
ODE. These variables need no communication whatsoever, in absence of
flux.  Thus, we can re-write the above loop as
\begin{verbatim}
  for (it = 0 to maxit) {
    evolve sources for dt/2
    evolve fluxes for dt
    update ghost zones for all grid functions which have a flux
    evolve sources for dt/2
  }
\end{verbatim}
In practice, this algorithm allows us to reduce our communication
cost, resulting in excellent scaling, in addition to excellent single
processor performance.

\begin{figure}
\label{fig:nt_scal}
\hspace{1.5cm}\epsfysize=8cm
\epsfbox{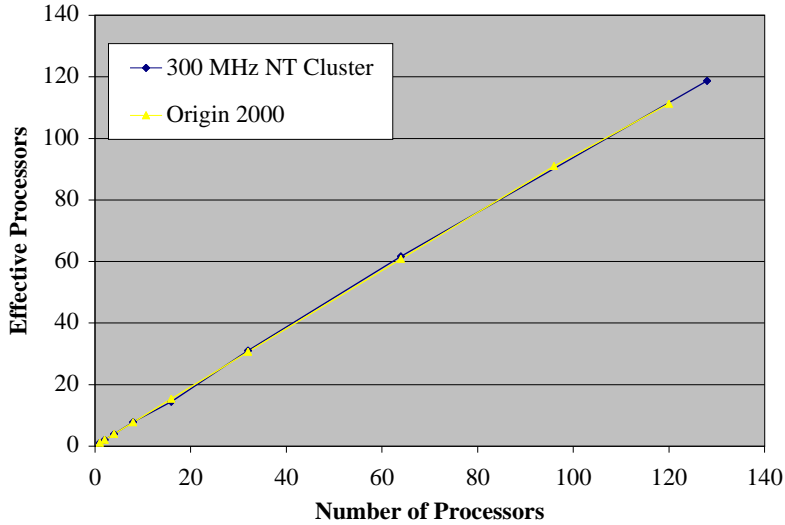}
\caption{We show scaling of the Cactus code on two very different
architectures:  an SGI/Cray Origin 2000 DSM architecture with 128
processors, and a cluster of 300Mhz Compaq workstations running
Windows NT.  The data is obtained with all "thorns"
that are needed to evolve a gravitational wave packet using the vacuum Einstein
equation.}
\end{figure}

\begin{figure}
\label{fig:cactus_scal}
\hspace{1.5cm}\epsfysize=8cm \epsfbox{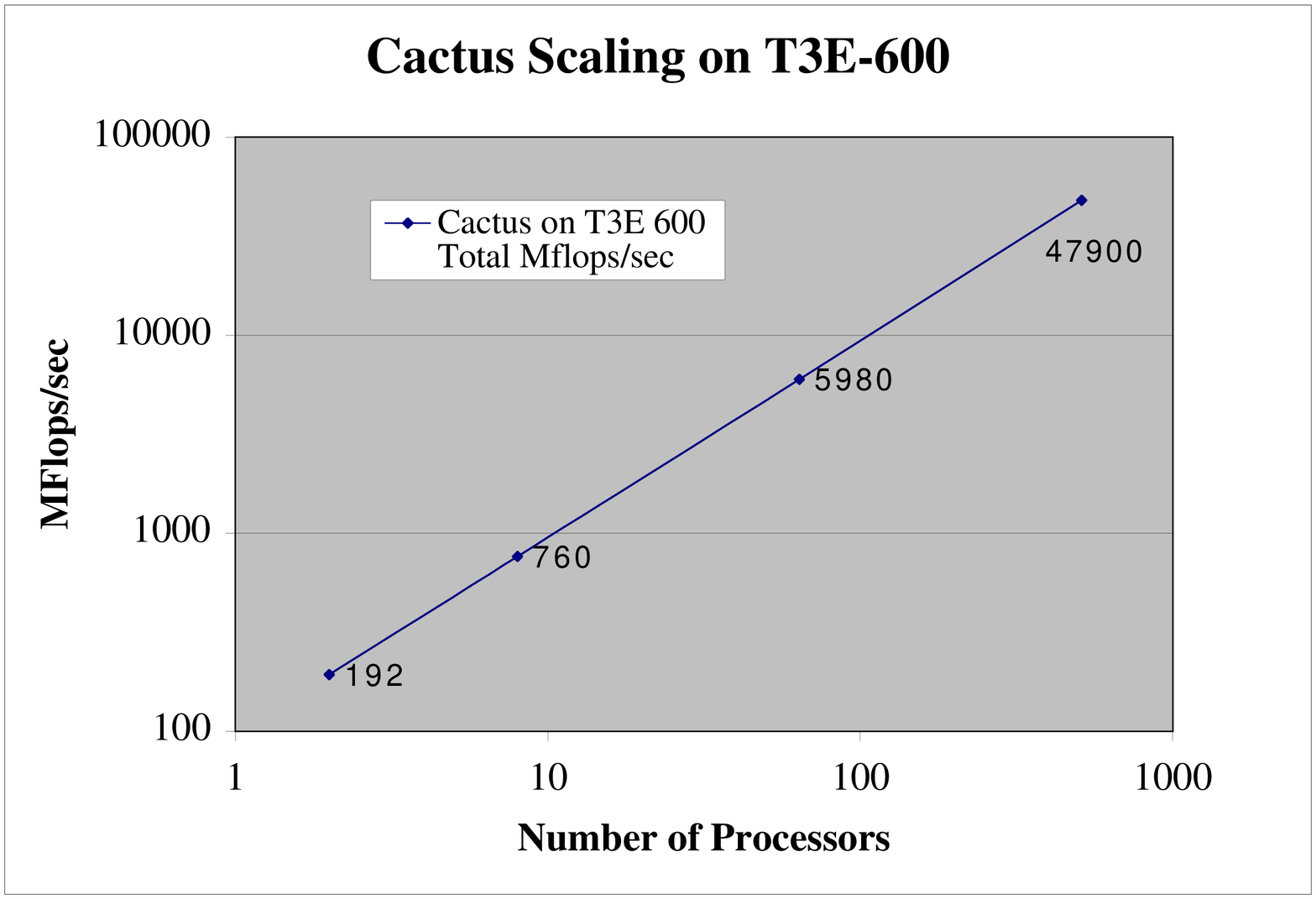}
\caption{We show the scaling of the Cactus code on the Cray T3E-600,
giving the total Mflops/sec as a function of the number of processors.
Results are shown for single precision calculations, and include
calculations performed on ghost zones.  The grid size per processor is
kept constant as the number of processors in increased (so the total
problem size scales with the size the machine).  The performance data
is obtained for an evolution using the Einstein equations with the
hydrodynamic source terms, and the relativistic hydrodynamics
equations in high resolution shock capturing treatment as described in
the paper.  The inclusion of the hydrodynamics does not change the
performance in any substantial manner.  }
\end{figure}

These techniques have enabled ``Cactus'' to be a highly portable and
efficient code for numerical relativity and astrophysics.  It has been
tested and performed very well on three very different parallel architectures:
the SGI Origin 2000 system, the SGI/Cray T3E system, and a cluster of
128 NT workstations, developed at NCSA, running Pentium II processors.
In Fig.~\ref{fig:cactus_scal} we show scaling results achieved on a
Cray T3E-600, and in Fig.~\ref{fig:nt_scal} we show the scaling
results achieved on both the Origin 2000 and the NT cluster.
Presently, the SGI Origin, with 250MHz R10000 processors, has more
than three times the per processor performance of the 300MHz Pentium II in
the NT cluster, but the trend is very encouraging.  These results
indicate that codes like this can be run efficiently in parallel on a
wide variety of machines.

Finally, we have recently tested the code on a 1024 node T3E-1200
(provided for the NASA Neutron Star Grand Challenge
Project~\cite{Nasa98} for performance tests), achieving 142GFlops and
linear scaling up to 1024 nodes.  The version of the code tested is
the so-called GR3D code developed for the NASA Grand Challenge
Project.  GR3D is a version of ``Cactus'' code for spacetime evolution
coupled to a Riemann solver based relativistic hydrodynamic code (MAHC
HYPERBOLIC\_HYDRO) that has recently been released (available at
http://wugrav.wustl.edu/Codes/GR3D/).
Performance information for the ``Cactus'' code will also be kept
up to date at
http://cactus.aei-potsdam.mpg.de.

In the following we give a summary of the performance test.  The test
was carried out with the released version without special tuning for
this 1024 node machine.  We note that the full set of the Einstein
equations with the perfect fluid source, as solved in this code,
involved a large number of 3D arrays.  The huge number of grid points
used (644 x 644 x 1284, or 500 x 500 x 996 respectively for 32 and 64
bits) is made possible by reduced memory footprint of the code.

\noindent {\em \underline{Machine Configuration}}: 1024 node T3E-1200 with
512MB per node

\noindent {\em \underline{Date tested}}: May 10, 1998

\begin{verbatim}
                                  32 bit             64 bit
--------------------------------------------------------------------
Grid Size per Processor           84x84x84           66x66x66
Processor topology                8 x8 x16           8 x8 x16
Total Grid Size                   644 x 644 x 1284   500 x 500 x 996
Single Proc MFlop/sec             144.35             118.33
Aggregate GFlop/sec               142.2              115.8
Scaling efficiency                96.2\%              95.6\%
--------------------------------------------------------------------
\end{verbatim}

Previously, the largest production simulations in 3D numerical
relativity have been limited to about $300^{3}$ or less, and applied
to distorted 3D black hole
systems~\cite{Camarda97b,Camarda97c,Allen97a,Allen98a}.  When such
large machines as the test T3E described above are made available for
routine production simulations, we expect to further improve the
results of the such black hole simulations, and perform more general
3D calculations involving distorted rotating black holes, coalescences
of neutron stars, gravitational waves, as well as other interesting
problems in general relativistic astrophysics.

\subsection{Collaborative Infrastructure}

While ``Cactus'' is our third generation 3D numerical relativity code,
it is our first generation of code in which we paid special attention
to the difficult software engineering problem of collaborative code
development, maintenance and management.  Our earlier generations of
3D numerical relativity codes (the so-called ``G'' and ``H'' codes,
described in~\cite{Anninos94c,Anninos94d}) involving about a dozen
researchers in the Potsdam/NCSA/Wash U collaboration, have made us
keenly aware of the issues and difficulties involved in distributed
collaborative code development.  For the ``Cactus'' code, a collaborative
infrastructure has been essential.  As of this writing, several dozen
collaborators at 7 institutions are using the code for various
research projects, and we aim at further making it a truly community
code for the investigation of general relativistic astrophysics.

``Cactus'' is hence designed to minimize barriers to the community
development and use of the code, including the complexity associated
with both the code itself and the networked supercomputer
environments in which simulations and data analyses are performed.
This complexity is particularly noticeable in large multidisciplinary
simulations such as ours, because of the range of disciplines that
must contribute to code development (relativity, hydrodynamics,
astrophysics, numerics, and computer science) and because of the
geographical distribution of the people and computer resources
involved in simulation and data analysis.

The collaborative technologies that we are developing within Cactus include:

\noindent $\bullet$ {\em A modular code structure and associated code 
development tools}.  Cactus defines coding rules that allow one, with 
only a working knowledge of Fortran or C, to write new code modules 
that are easily plugged in as ``thorns'' to the main Cactus code (the 
``flesh'').  The ``flesh'' contains basic computational infrastructure 
needed to develop and connect parallel modules into a working code.  
All told, the ``flesh'' is thousands of lines of highly optimized C 
and Fortran, not counting some utility libraries, makefile, and perl 
scripts.  It allows one to plug in not only different physics modules, 
such as the basic Einstein solver, other formulations of the 
equations, analysis routines, etc., but also different parallel domain 
decomposition modules, I/O modules, utilities, elliptic solvers, and 
so forth.  A thorn may be any code that the user wants, in order to 
provide different initial data, different matter fields, different 
gauge conditions, visualization modules, etc.  Thorns need not have 
anything to do with relativity: the flesh could be used as basic 
infrastructure for any set of PDE's, from Newtonian hydrodynamics 
equations to Yang Mills equations, that are coded as thorns.  The user 
inserts the hook to their thorn into the flesh code in a way that the 
thorn will not be compiled unless it is designated to be active.  We 
have developed a makefile and perl-based thorn management system that, 
through the use of preprocessor macros that are appropriately expanded 
to the arguments of the flesh and additional arguments defined by each 
thorn, is able to create a code which can configure itself to have a 
variety of different modules.  This ensures that {\em only} those 
variables needed for a particular simulation are actually used, and 
that no conflicts can be created in subroutine argument calling lists.

\noindent $\bullet$
{\em A consistency test suite library.}  An increased number of thorns
makes the code more attractive to its community but also increases the
risk of incompatibilities.  Hence, we provide technology that allows
each developer to create a test/validation suite for their own thorn.
These tests are run prior to any check in of code to the repository,
ensuring that new code reproduces results consistent with previous one,
and hence cannot compromise the work of other developers relying on
a given thorn.

So, how does a user use the code?  A detailed user guide will be 
available with the code when it is released during 1999 (see 
http://cactus.aei-potsdam.mpg.de), but in a nutshell, one specifies 
which physics modules, and which computational/parallelism modules, 
are desired in a configuration file, and makes the code on the desired 
architecture, which can be any one of a number of machines from 
SGI/Cray Origin or T3E, Dec Alpha, Linux workstations or clusters, NT 
clusters, and others.  The make system automatically detects the 
architecture and configures the code appropriately.  Control of run 
parameters is then provided through an input file.  Additional modules 
can be selected from a large community-developed library, or new 
modules may be written and used in conjunction with the pre-developed 
modules.

Our experiences with Cactus up to now suggest that these techniques
are effective.  It allows a code of many tens of thousands of lines,
but with a compact flesh that is possible to maintain despite the
large number of people contributing to it.  The common code base has
enhanced the collaborative process, having important and beneficial
effects on the flow of ideas between remote groups.  This flexible,
open code architecture allows, for example, a relativity expert to
contribute to the code without knowing the details of, say, the
computational layers (e.g., message passing or AMR libraries) or other
components (e.g., hydrodynamics).  This is an area that we plan to
invest more effort into in the coming few years.

\subsection{Adaptive Mesh Refinement}

3D simulations of Einstein's equations are very demanding
computationally.  In this section we outline the computational needs,
and techniques designed to reduce them.  We will need to resolve waves
with wavelengths of order $5M$ or less, where $M$ is the mass of the
BH or the neutron star.  Although for Schwarzschild black holes, the
fundamental $\ell=2$ quasinormal mode wavelength is $16.8M$, higher
modes, such as $\ell=4$ and above, have wavelengths of $8M$ and below.
The BH itself has a radius of $2M$.  More important, for very rapidly
rotating Kerr BH's, which are expected to be formed in realistic
astrophysical BH coalescence, the modes are shifted down to
significantly shorter wavelengths\cite{Flanagan97a,Flanagan97b}.  As
we need at least 20 grid zones to resolve a single wavelength, we can
conservatively estimate a required grid resolution of about $\Delta x
= \Delta y =
\Delta z \approx 0.2M$.  For simulations of time scales of order $t
\propto 10^{2}-10^{3}M$, which will be required to follow coalescence,
the outer boundary will probably be placed at a distance of roughly $R
\propto 100M$ from the coalescence, requiring a Cartesian simulation
domain of about $200M$ across.  This leads to about $10^{3}$ grid
zones in each dimension, or about $10^{9}$ grid zones in total.  As 3D
codes to solve the full Einstein equations have typically 100
variables to be stored at each location, and simulations are performed
in double precision arithmetic, this leads to a memory requirement of
order 1000 Gbytes!  (In fairness to some groups that use spectral
methods instead of finite differences (e.g., the Meudon group), we
should point out highly accurate 3D simulations can now be achieved on
problems that are well suited to such techniques, using much less
memory!~\cite{Bonazzola98a}).

The largest supercomputers available to scientific research
communities today have only about $\frac{1}{20}$ of this capacity, and
machines with such capacity will not be routinely available for some
years.  Furthermore, if one needs to double the resolution in each
direction for a more refined simulation, the memory requirements
increase by an order of magnitude.  Although such estimates will vary,
depending on the ultimate effectiveness of inner or outer boundary
treatments, gauge conditions, etc., they indicate that barring some
unforeseen simplification, some form of adaptive mesh refinement (AMR)
that places resolution only where it is required is not only
desirable, but essential.  The basic idea of AMR is to use some set of
criteria to evaluate the quality of the solution on the present time
step.  If there are regions that require more resolution, then data
are interpolated onto a finer grid in those regions; if less
resolution is required, grid points are destroyed.  Then the evolution
proceeds to the next time step on this hierarchy of grids, where the
process is repeated.  These rough ideas have been refined and applied
in many applications now in computational science.

There are several efforts ongoing in AMR for relativity.  Choptuik was
the early pioneer in this area, developing a 1D AMR system to handle
the resolution requirements needed to follow scalar field collapse to
a BH\cite{Choptuik89}.  As an initially regular distribution of scalar
field collapses, it will require more and more resolution as its
density builds up.  The grid density required to resolve the initial
distribution may not even see the final BH. Further, as pulses of
radiation propagate back out from the origin, they, too may have to be
resolved in regions where there was previously a coarse grid.
Choptuik's AMR system, built on early work of Berger and
Oliger\cite{Berger84}, was able to track dynamically features that
develop, enabling him to discover and accurately measure BH critical
phenomena that have now become so widely studied\cite{Choptuik93}.

Based on this success and others, and on the general considerations 
discussed above, full 3D AMR systems are under development to handle 
the much greater needs of solving the full set of 3D Einstein 
equations.  A large collaboration, begun by the NSF Black Hole Grand 
Challenge Alliance, has been developing a system for distributing 
computing on large parallel machines, called Distributed Adapted Grid 
Hierarchies, or DAGH. DAGH was developed to provide MPI-based 
parallelism for the kinds of computations needed for many PDE solvers, 
and it also provides a framework for parallel AMR. It is one of the 
major computational science accomplishments to come out of the 
Alliance.  Developed by Manish Parashar and Jim Browne, in 
collaboration with many subgroups within and without the Alliance, it 
is now being applied to many problems in science and engineering.  One 
can find information about DAGH online at 
http://www.cs.utexas.edu/users/dagh/.

At least two other 3D software environments for AMR have been
developed for relativity: one is called HLL, or Hierarchical Linked
Lists, developed by Lee Wild and Bernard Schutz\cite{Wild98a};
another, called BAM, was the first AMR application in 3D relativity
developed by Br{\"u}gmann~\cite{Bruegmann96}.  The HLL system has
recently been applied to the test problem of the Zerilli equation
(discussed above) describing perturbations of black
holes\cite{Papadopoulos98a}.  This nearly 30 year old linear equation
is still providing a powerful model for studying BH collisions, and it
is also being used as a model problem for 3D AMR. In this work, the 1D
Zerilli equation is recast as a 3D equation in cartesian coordinates,
and evolved within the AMR system provided by HLL. Even though the 3D
Zerilli equation is a single linear equation, it is quite demanding in
terms of resolution requirements, and without AMR it is extremely
difficult to resolve both the initial pulse of radiation, the blue
shifting of waves as they approach the horizon, and the scattering of
radiation, including the normal modes, far from the hole.

\section{Summary}
In this article we have attempted to review the essential mathematical
and computational elements needed for a full scale numerical
relativity code that can treat a variety of problems in relativistic
astrophysics and gravitation.  Various formulations of the Einstein
equations for evolving spacelike time slices, techniques for providing
initial data, the basic ideas of gauge conditions, several important
analysis tools for discovering the physics contained in a simulation,
and numerical algorithms for each of these items have been reviewed.
Unfortunately, we have only been able to cover the basics of such a
program, and in addition many promising alternative approaches have
necessarily been left out.

As one can see, the solution to a single problem in numerical
relativity requires a huge range of computational and mathematical
techniques.  It is truly a large scale effort, involving experts in
computer and computational science, mathematical relativity,
astrophysics, and so on.  For these reasons, aided by collaborations
such as the NSF Black Hole Grand Challenge Alliance and the
NCSA/Potsdam/WashU collaboration, there has been a great focusing of
effort over the last years.

A natural byproduct of this focusing has been the development of
codes that are used and extended by large groups.  A code must have a
large arsenal of modules at its disposal: different initial data sets,
gauge conditions, horizon finders, slicing conditions, waveform
extraction, elliptic equation solvers, AMR systems, boundary modules,
different evolution modules, etc.  Furthermore, these codes must run
efficiently on the most advanced supercomputers available.  Clearly,
the development of such a sophisticated code is beyond any single
person or group.  In fact, it is beyond the capability of a single
community!  Different research communities, from computer science,
physics, and astrophysics, must work together to develop such a code.

As an example of such a project, we described briefly the ``Cactus'' 
code, developed by a large international collaboration\cite{Bona98b}.  
This code is an outgrowth of the last 5 years of 3D numerical 
relativity development primarily at NCSA/Potsdam/WashU, and builds 
heavily on the experience gained in developing the so-called ``G'' and 
``H'' codes~\cite{Anninos94c,Anninos94d,Bona98b}.  Presently, it is 
being developed collaboratively by these groups in collaboration with 
groups at Palma, Valencia, Physical Research Laboratory in India, and 
computational science groups at U. of Illinois, and Argonne National 
Lab.

The code has a very modular structure, allowing different physics,
analysis, and computational science modules to be plugged in.  In
fact, versions of essentially all the modules listed above are already
developed for the code.  For example, several formulations of
Einstein's equations, including the ADM formalism and the Bona-Mass\'o
hyperbolic formulation, can be chosen as input parameters, as can
different gauge conditions, horizon finders, hydrodynamics evolvers,
etc.  It is being tested on BH spacetimes, such as those described
above, as well as on pure wave spacetimes, self-gravitating scalar
fields and hydrodynamics.  It has also been designed to connect to
DAGH ultimately for parallel AMR.

This code was also designed as a community code.  After first
developing and testing it within our rather large community of
collaborators, it will be made available with full documentation via
a public ftp server maintained at AEI and a mirroring site at
WashU.  By having an entire research
community using and contributing to such a code, we hope to
accelerate the maturation of numerical relativity.  Information about
the code is available online, and can be accessed at
http://cactus.aei-potsdam.mpg.de.

{\bf Acknowledgments} {It is a pleasure to acknowledge many friends
and colleagues who have contributed to the work described in this
article, some of which was derived from papers we have written
together.  The Cactus code was originally started by Joan Mass\'o and
Paul Walker at AEI, and the MAHC general relativistic hydrodynamic
module was started by Mark Miller at Wash U; these were then opened up
to development by our entire research groups at Potsdam, Washington
University, and NCSA (the NCSA/Potsdam/Wash U collaboration), and
elsewhere, notably the University of the Balearic Islands in Spain,
the University of Valencia, Argonne National Lab, and Physical
Research Lab (PRL) in India.  Without the contributions from people at
all these institutions, the work described here would not have been
possible.  Thanks to Tom Goodale and Ed Evans for carefully reading
the manuscript and suggesting improvements, to Tom Clune for
performance tuning and scaling results for the T3E, and to staff at
NCSA for helping study the performance of the code on the Origin 2000
and NT workstations.  This work has been supported by AEI, NCSA, NSF
grant No.  PHY-96-00507, NASA HPCC/ESS Grand Challenge Applications
Grant No.  NCCS5-153, NSF MRAC Allocation Grant No.~MCA93S025, and
Hong Kong RGC Grant CUHK 4189/97P.}


\end{document}